\documentclass[aps,pre,twocolumn,groupedaddress,showpacs,showkeys]{revtex4}

\usepackage{graphicx}
\usepackage{dcolumn}
\usepackage{bm}
\usepackage{amssymb}

\begin{document}


\title{An approach to chaotic synchronization}


\author{Alexander~E.~Hramov}
\email{aeh@cas.ssu.runnet.ru}
\author{Alexey~A.~Koronovskii}
\email{alkor@cas.ssu.runnet.ru}
\affiliation{Department of Nonlinear Processes, Saratov State
University, Astrakhanskaya, 83, Saratov, 410012, Russia}



\date{\today}

\begin{abstract}
This paper deals with the chaotic oscillator synchronization. A
new approach to the synchronization of chaotic oscillators has
been proposed. This approach is based on the analysis of different
time scales in the time series generated by the coupled chaotic
oscillators. It has been shown that complete synchronization,
phase synchronization, lag synchronization and generalized
synchronization are the particular cases of the synchronized
behavior called as ``time--scale synchronization''. The
quantitative measure of chaotic oscillator synchronous behavior
has been proposed. This approach has been applied for the coupled
R\"ossler systems and two coupled Chua's circuits.
\end{abstract}


\maketitle


{\bf Synchronization of chaotic oscillators is one of the
fundamental phenomena of nonlinear dynamics. There are several
different types of synchronization of coupled chaotic oscillators
which have been described theoretically and observed
experimentally. In this paper we propose a new approach to the
synchronization of two coupled chaotic oscillators based on the
consideration of oscillators' time scale dynamics. We have shown
that synchronized behavior of time scales should be considered as
a new type of synchronization called as \textit{``time scale
synchronization''} and other types of chaotic synchronization are
the particular cases of time scale synchronization.}

\section*{Introduction}
\label{intro} Synchronization of chaotic oscillators is one of the
fundamental phenomena of nonlinear dynamics. It takes place in
many
physical~\cite{Parlitz:1996_PhaseSynchroExperimental,%
Tang:1998_PhaseSynchroLasers,%
Allaria:2001_PhaseSynchroLaser,Ticos:2000_PlasmaDischarge,%
Rosa:2000_PlasmaDischarge,aeh:2003_SynchroDistrSyst} and
biological~\cite{Tass:1998_NeuroSynchro,Anishchenko:2000_humanSynchro,%
Prokhorov:2003_HumanSynchroPRE} processes. It seems to play an
important role in the ability of biological oscillators, such as
neurons, to act
cooperatively~\cite{Elson:1998_NeronSynchro,Rulkov:2002_2DMap,%
Tass:2003_NeuroSynchro}. Chaotic synchronization can also be used
for secret signal
transmission~\cite{Murali:1993_SignalTransmission,Chua:1997_Criptography}.
There are several different types of synchronization of coupled
chaotic oscillators which have been described theoretically and
observed experimentally~\cite{Pikovsky:2002_SynhroBook,%
Anshchenko:2001_SynhroBook,Pikovsky:2000_SynchroReview,%
Anishchenko:2002_SynchroEng}. These are the \textit{complete
synchronization} (CS) \cite{Pecora:1990_ChaosSynchro,%
Pecora:1991_ChaosSynchro}, \textit{lag synchronization} (LS)
\cite{Rosenblum:1997_LagSynchro,Taherion:1999_LagSynchro},
\textit{generalized synchronization} (GS)
\cite{Rulkov:1995_GeneralSynchro,Kocarev:1996_GS} and
\textit{phase synchronization} (PS)
\cite{Pikovsky:2002_SynhroBook,Anshchenko:2001_SynhroBook}. In
this paper we propose a new approach to the synchronization of two
coupled chaotic oscillators based on the consideration of
oscillators' time scale dynamics. We have shown that synchronized
behavior of time scales should be considered as a new type of
synchronization called as \textit{``time scale synchronization''}
(TSS) and CS, LS, PS, GS are the particular cases of TSS.

The \textit{complete synchronization} (CS) implies coincidence of
states of coupled oscillators
$\mathbf{x}_1(t)\cong\mathbf{x}_2(t)$, the difference between
state vectors of coupled systems converges to zero in the limit
$t\rightarrow\infty$,
\cite{Pecora:1990_ChaosSynchro,%
Pecora:1991_ChaosSynchro,Murali:1994_SynchroIdenticalSyst,%
Murali:1993_SignalTransmission}. It appears when interacting
systems are identical. If the parameters of coupled chaotic
oscillators slightly mismathch, the state vectors are close
$|\mathbf{x}_1(t)-\mathbf{x}_2(t)|\approx 0$, but differ from each
other. Another type of synchronized behavior of coupled chaotic
oscillators wihth slightly mismatched parameters is the
\textit{lag synchronization} (LS), when shifted in time, the state
vectors coincide with each other,
$\mathbf{x}_1(t+\tau)=\mathbf{x}_2(t)$. When the coupling between
oscillator increases the time lag $\tau$ decreases and the
synchronization regime tends to be CS
one~\cite{Rosenblum:1997_LagSynchro,Zhigang:2000_GSversusPS,%
Taherion:1999_LagSynchro}. The \textit{generalized
synchronization} (GS)
\cite{Rulkov:1995_GeneralSynchro,Kocarev:1996_GS,%
Pyragas:1996_WeakAndStrongSynchro} introduced for drive--responce
systems, means that there is some functional relation between
coupled chaotic oscillators, i.e.
$\mathbf{x}_2(t)=\mathbf{F}[\mathbf{x}_1(t)]$.

Finally, it is necessary to mention the \textit{phase
synchronization} (PS) regime. To describe the phase
synchronization the instantaneous phase $\phi(t)$ of a chaotic
continuous time series is usually
introduced \cite{Pikovsky:2000_SynchroReview,Anishchenko:2002_SynchroEng,%
Pikovsky:2002_SynhroBook,Anshchenko:2001_SynhroBook,%
Rosenblum:1996_PhaseSynchro,Osipov:1997_PhaseSynchro}. The phase
synchronization means the entrainment of phases of chaotic
signals, whereas their amplitudes remain chaotic and uncorrelated.

All synchronization types mentioned above are associated with each
other (see for detail~\cite{Parlitz:1996_PhaseSynchroExperimental,%
Rulkov:1995_GeneralSynchro,Zhigang:2000_GSversusPS}), but the
relationship between them is not completely clarified yet. For
each type of synchronization there are their own ways to detect
the synchronized behavior of coupled chaotic oscillators. The
complete synchronization can be displayed by means of comparison
of system state vectors $\mathbf{x}_1(t)$ and $\mathbf{x}_2(t)$,
whereas the lag synchronization can be determined by means of the
similarity function~\cite{Rosenblum:1997_LagSynchro}. The case of
the generalized synchronization is more intricate because the
functional relation $\mathbf{F}[\cdot]$ can be very complicated,
but there are several methods to detect the synchronized behavior
of coupled chaotic oscillators, such as the auxiliary system
approach~\cite{Rulkov:1996_AuxiliarySystem} or the method of
nearest
neighbors~\cite{Rulkov:1995_GeneralSynchro,Pecora:1995_statistics}.

Finally, the phase synchronization of two coupled chaotic
oscillators occurs if the difference between the instantaneous
phases $\phi_{1,2}(t)$ of chaotic signals $\mathbf{x}_{1,2}(t)$ is
bounded by some constant
\begin{equation}
|\phi_{1}(t)-\phi_{2}(t)|<\mathrm{const}. \label{eq:PhaseLocking}
\end{equation}
It is possible to define a mean frequency of chaotic signal
\begin{equation}
\bar{\Omega}=\lim\limits_{t\rightarrow\infty}\frac{\phi(t)}{t}=
\langle\dot{\phi}(t)\rangle, \label{eq:MeanFrequency}
\end{equation}
which should be the same for both coupled chaotic systems, i.e.,
the phase locking leads to the frequency entrainment. It is
important to notice, to obtain correct results the mean frequency
$\bar{\Omega}$ of chaotic signal $\mathbf{x}(t)$ should coincide
with the main frequency $\Omega_0=2\pi f_0$ of the Fourier
spectrum (for detail, see~\cite{Anishchenko:2004_ChaosSynchro}).

In this paper we propose a new approach to the synchronization of
two coupled chaotic oscillators. The main idea of this approach
consists in the analysis of the system behavior on different time
scales that allows us to consider different cases of
synchronization from the universal positions. The new type of
synchronous behavior (so--called time scale synchronization (TSS))
has been introduced.

The structure of this paper is the following. In
section~\ref{Sct:WVTTrans} we discuss the method of the time
scales $s$ and associated with them phases $\phi_s(t)$ definition
by means of the continuous wavelet transform. The concept of time
scale synchronization is given in section~\ref{Sct:TSSynchro}.
Section~\ref{Sct:IllPhase} deals with the synchronization of two
mutually coupled R\"ossler systems with funnel attractors. We
demonstrate the efficiency of our method for such cases and
discuss the correlation between PS, LS, CS and TSS.
Section~\ref{Sct:Chua} deals with the time scale synchronization
of two coupled chaotic Chua's circuits. In
section~\ref{Sct:GSSynchro} we consider application of our method
for the unidirectional coupled R\"ossler systems when the
generalized synchronization is observed. The quantitative measure
of synchronization is described in section~\ref{Sct:Measure}. The
final conclusion is presented in section~\ref{Sct:Conclusion}.

\section{Continuous wavelet transform and time scales dynamics}
\label{Sct:WVTTrans}

The continuous wavelet transform~\cite{alkor:2003_WVTBookEng,%
Daubechies:1992_WVTBook,Kaiser:1994_Wvt,Torresani:1995_WVT} is the
powerful tool for the analysis of nonlinear dynamical system
behavior. In particular, the continuous wavelet analysis has been
used for the detection of synchronization of chaotic oscillations
in the
brain~\cite{Lachaux:2000_WVTSynchro,Lachaux:2002_BrainCoherence,%
Quyen:2001_WVTvsHilbert} and chaotic laser
array~\cite{DeShazer:2001_WVT_LaserArray}. It has also been used
to detect the main frequency of the oscillations in nephron
autoregulation~\cite{Sosnovtseva:2002_Wvt}. We propose to analyze
the dynamics of coupled chaotic oscillators using the
consideration of system behavior on different time scales $s$ each
of them is characterized by means of its own phase $\phi_s(t)$,
respectively. So, in order to define {\it the continuous set of
instantaneous phases} $\phi_s(t)$ the continuous wavelet transform
is the convenient mathematical tool.

Let us consider continuous wavelet transform of chaotic time
series $x(t)$
\begin{equation}
W(s,t_0)=\int\limits_{-\infty}^{+\infty}x(t)\psi^*_{s,t_0}(t)\,dt,
\label{eq:WvtTrans}
\end{equation}
where $\psi_{s,t_0}(t)$ is the wavelet--function related to the
mother--wavelet $\psi_{0}(t)$ as
\begin{equation}
\psi_{s,t_0}(t)=\frac{1}{\sqrt{s}}\psi\left(\frac{t-t_0}{s}\right).
\label{eq:Wvt}
\end{equation}
The time scale $s$ corresponds to the width of the wavelet
function $\psi_{s,t_0}(t)$, and $t_0$ is shift of wavelet along
the time axis, the symbol ``$*$'' in~(\ref{eq:WvtTrans}) denotes
complex conjugation. It should be noted that the time scale $s$ is
usually used instead of the frequency $f$ of Fourier
transformation and can be considered as the quantity inversed to
it.

The Morlet--wavelet~\cite{Grossman:1984_Morlet}
\begin{equation}
\psi_0(\eta)=\frac{1}{\sqrt[4]{\pi}}\exp(j\Omega_0\eta)\exp\left(\frac{-\eta^2}{2}\right)
\label{eq:Morlet}
\end{equation}
has been used as a mother--wavelet function. The choice of
parameter value $\Omega_0=2\pi$ provides the relation ${s=1/f}$
between the time scale $s$ of wavelet transform and frequency $f$
of Fourier transformation.

The wavelet surface
\begin{equation}
W(s,t_0)=|W(s,t_0)|e^{j\phi_s(t_0)} \label{eq:WVT_Phase}
\end{equation}
describes the system's dynamics on every time scale $s$ at the
moment of time $t_0$. The value of $|W(s,t_0)|$ indicates the
presence and intensity of the time scale $s$ mode in the time
series $x(t)$ at the moment of time $t_0$. It is possible to
consider the quantitie
\begin{equation}
\langle E(s)\rangle=\int|W(s,t_0)|^2\,dt_0, \label{eq:IntEnergy}
\end{equation}
which is the integral energy distribution on time scales,
respectively. At the same time, the phase $\phi_s(t)=\arg\,W(s,t)$
is naturally introduced for every time scale $s$. In other words,
$\phi_s(t)$ is continuous function of time $t$ and time scale $s$.

\section{Time scale synchronization}
\label{Sct:TSSynchro}

Using the continuous wavelet transform we have introduced the
continuous set of time scales $s$ and associated with them
instantaneous phases $\phi_s(t)$ (see section~\ref{Sct:WVTTrans}).
It means that it is possible to describe the behavior of each time
scale $s$ by means of its own phase $\phi_s(t)$. Let us consider
the dynamics of two coupled oscillators. If in the time series
$\mathbf{x}_{1,2}(t)$ generated by these systems there is time
scale range $s_m\leq s\leq s_b$ for time scales $s$ from which the
phase locking condition
\begin{equation}
|\phi_{s1}(t)-\phi_{s2}(t)|<\mathrm{const}
\label{eq:SPhaseLocking}
\end{equation}
is satisfied and the part of the wavelet spectrum energy being
fallen on this range is not equal to zero
\begin{equation}
E_{snhr}=\int\limits_{s_m}^{s_b}\langle E(s)\rangle\,ds>0,
\label{eq:SynchroEnergy}
\end{equation}
we say that \textit{time scale synchronization} (TSS) between
oscillators takes place.

It is obvious that the classical synchronization of coupled
periodical oscillators is equal to TSS because in this case all
time scales are synchronized according to the time scale $s$,
instantaneous phase $\phi_s(t)$ and TSS definitions. The case of
chaotic oscillations is more complicated. Nevertheless, as we will
show further, if two chaotic oscillators demonstrate any type of
synchronized behavior mentioned above (CS, LS, PS or GS), in the
time series $\mathbf{x}_{1,2}(t)$ generated by these systems there
are time scales $s$ necessarily correlated with each other for
which the phase locking condition~(\ref{eq:SPhaseLocking}) is
satisfied. Therefore, the time scale synchronization is also
realized. In other words, CS, LS, PS and GS are the particular
cases of the time--scale synchronization. To detect time scale
synchronization one can examine the
condition~(\ref{eq:SPhaseLocking}) which should be satisfied for
synchronized time scales.

Let us consider several examples of coupled oscillators
demonstrating different types of chaotic synchronization (PS, LS,
GS). We will show that if any type of synchronous behavior is
observed the time scale synchronization is detected, too. So, TSS
is general case of synchronization and CS, LS, PS, GS are the
particular cases of TSS.

\section{Example I. Time scale synchronization of two
R\"ossler systems: from phase to lag synchronization}
\label{Sct:IllPhase}

Let us start our consideration with two mutually coupled R\"ossler
systems with slightly mismatched parameters. For this system it is
impossible to correctly introduce the instantaneous phase
$\phi(t)$ of chaotic signal $\mathbf{x}(t)$. It is clear, that for
such cases the traditional methods of the phase synchronization
detecting fail and it is necessary to use the other techniques,
e.g., indirect
measurements~\cite{Rosenblum:2002_FrequencyMeasurement}. On the
contrary, our approach allows easily to detect the time scale
synchronization between chaotic oscillators.

\begin{figure}
 \centerline{\includegraphics*[scale=0.4]{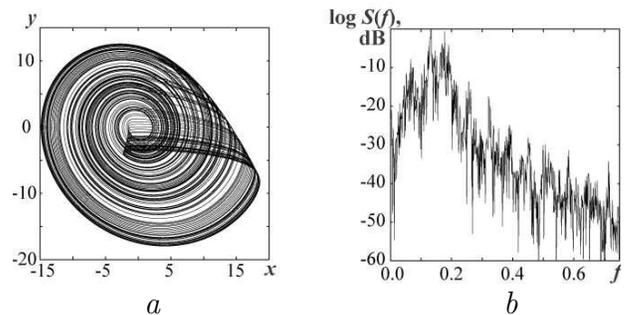}}
\centerline{\large\quad\textit{a}\qquad\qquad\qquad\qquad\qquad\quad\textit{b}}
\caption{Phase picture and power spectrum of the first R\"ossler
system~(\ref{eq:FunnelRoessler}) oscillations. Coupling parameter
$\varepsilon$ is equal to zero \label{fgr:FunnelRoessler}}
\end{figure}

To illustrate it we consider two non--identical coupled R\"ossler
systems with funnel attractors (Fig.~\ref{fgr:FunnelRoessler}):
\begin{equation}
\begin{array}{l}
\dot
x_{1,2}=-\omega_{1,2}y_{1,2}-z_{1,2}+\varepsilon(x_{2,1}-x_{1,2}),\\
\dot y_{1,2}=\omega_{1,2}x_{1,2}+ay_{1,2}+\varepsilon(y_{2,1}-y_{1,2}),\\
\dot z_{1,2}=p+z_{1,2}(x_{1,2}-c), \label{eq:FunnelRoessler}
\end{array}
\end{equation}
where $\varepsilon$ is a coupling parameter, $\omega_1=0.98$,
$\omega_2=1.03$. The control parameter values have been selected
by analogy with~\cite{Rosenblum:2002_FrequencyMeasurement} as
$a=0.22$, $p=0.1$, $c=8.5$. It is necessary to note that under
these control parameter values none of the methods mentioned above
permits to define phase of chaotic signal correctly in whole range
of coupling parameter $\varepsilon$ variation. Therefore, nobody
can determine by means of direct measurements whether the phase
synchronization regime takes place for several values of parameter
$\varepsilon$. On the contrary, our approach allows to detect TSS
synchronization between considered coupled oscillators easily for
all values of coupling parameter.

\begin{figure}
 \centerline{\includegraphics*[scale=0.5]{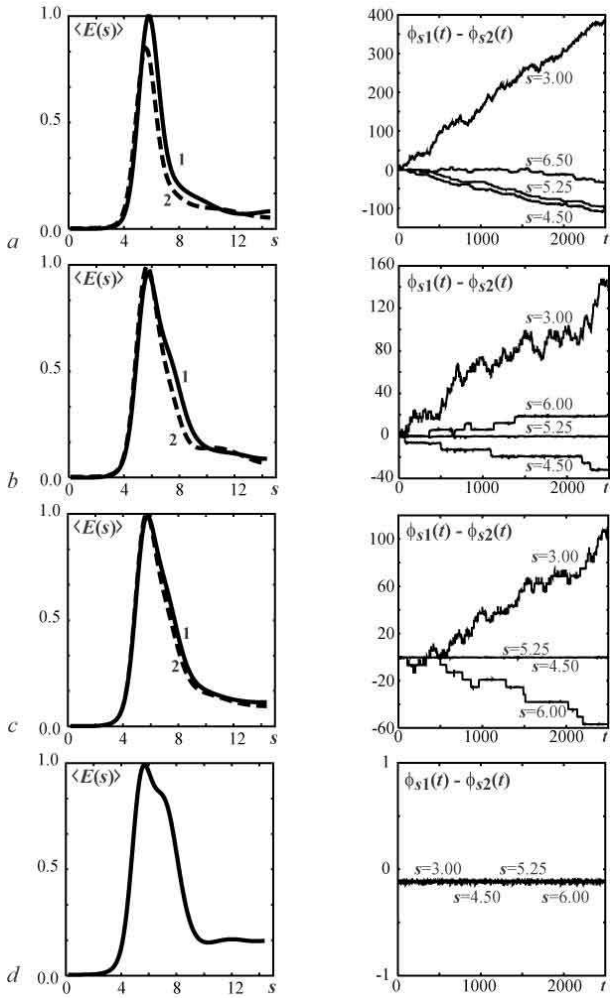}}
\caption{The normalized energy distribution in wavelet spectrum
$\langle E(s)\rangle$ for the first (the solid line denoted ``1'')
and the second (the dashed line denoted ``2'') R\"ossler
systems~(\ref{eq:FunnelRoessler}). The phase difference
$\phi_{s1}(t)-\phi_{s2}(t)$ for two coupled R\"ossler systems.
(\textit{a}) The value of coupling parameter has been selected as
$\varepsilon=0.025$. There is no phase synchronization between
systems; (\textit{b}) the value of coupling parameter has been
selected as $\varepsilon=0.05$. The time scales $s=5.25$ are
correlated with each other and the synchronization has been
observed; (\textit{c}) the value of coupling parameter has been
selected as $\varepsilon=0.07$; \textit{d}) The value of coupling
parameter has been selected as $\varepsilon=0.25$. The lag
synchronization has been observed, all time scales are
synchronized \label{fgr:FunRossK}}
\end{figure}

In~\cite{Rosenblum:2002_FrequencyMeasurement} it has been shown by
means of the indirect measurements that for the coupling parameter
value $\varepsilon=0.05$ the synchronization of two mutually
coupled R\"ossler systems~(\ref{eq:FunnelRoessler}) takes place.
Our approach based on the analysis of the dynamics of different
time scales $s$ gives analogous results. So, the behavior of the
phase difference $\phi_{s1}(t)-\phi_{s2}(t)$ for this case has
been presented in figure~\ref{fgr:FunRossK},\textit{b}. One can
see that the phase locking takes place for the time scales
$s=5.25$ which are characterized by the largest energy value in
the wavelet power spectra $\langle E(s)\rangle$ (see
Fig.~\ref{fgr:FunRossK},\textit{b}).

It is important to note that the phase difference
$\phi_{s1}(t)-\phi_{s2}(t)$ is also bounded on the time scales
close to $s=5.25$. So, we can say that the time scales $s=5.25$
(and close to them) of two oscillators are synchronized with each
other. At the same time the other time scales (e.g., $s=4.5, 6.0$
et. al.) remain uncorrelated. For such time scales the phase
locking has not been observed (see
Fig.~\ref{fgr:FunRossK},\textit{b}).

It is clear, the synchronization phenomenon is caused by the
existence of time scales $s$ in system dynamics correlated with
each other. It has been shown in~\cite{Rosenblum:1997_LagSynchro}
that there is certain relationship between PS, LS and CS for
chaotic oscillators with slightly mismatched parameters. With the
increase of coupling strength the systems undergo the transition
from unsynchronized chaotic oscillations to the phase
synchronization. With a further increase of coupling the lag
synchronization is observed. Further increasing of the coupling
parameter leads to the decreasing of the time lag and both systems
tend to have the complete synchronization regime.

Let us consider the dynamics of different time scales $s$ of two
nonidentical mutually coupled R\"ossler
systems~(\ref{eq:FunnelRoessler}) when the coupling parameter
value increases. If there is no phase synchronization between the
oscillators, then their dynamics remain uncorrelated for all time
scales $s$. Figure~\ref{fgr:FunRossK},\textit{a} illustrates the
dynamics of two coupled R\"ossler systems when the coupling
parameter $\varepsilon$ is small enough ($\varepsilon=0.025$). The
power spectra $\langle E(s)\rangle$ of wavelet transform for
R\"ossler systems differ from each other
(Fig.~\ref{fgr:FunRossK},\textit{a}), but the maximum values of
the energy correspond approximately to the same time scale $s$ in
both systems. It is clear, that the phase difference
$\phi_{s1}(t)-\phi_{s2}(t)$ is not bounded for almost all time
scales (see Fig.~\ref{fgr:FunRossK},\textit{a}). One can see that
the phase difference $\varphi_{s1}(t)-\varphi_{s2}(t)$ increases
for time scale $s=3.0$, but decreases for $s=4.5$. It means that
there should be the time scale $3<s^*<4.5$ the phase difference on
which remains bounded. This time scale $s^*$ plays a role of a
point separating the time scale areas with the phase difference
increasing and decreasing, respectively. In this case the measure
of time scales on which the phase difference remains bounded is
zero (therefore, the synchronized energy of wavelet power
spectra~(\ref{eq:SynchroEnergy}) is equal to zero) and we can not
say about the synchronous behavior of coupled chaotic oscillators
(see also section~\ref{Sct:Measure}).

As soon as any of the time scales of the first chaotic oscillator
becomes correlated with the other one of the second oscillator
(e.g., when the coupling parameter increases), the phase
synchronization occurs (see Fig.~\ref{fgr:FunRossK},\textit{b}).
The time scales $s$ characterized by the largest value of energy
in wavelet spectrum $\langle E(s)\rangle$ are more likely to
become correlated first. The other time scales remain uncorrelated
as before. The phase synchronization between chaotic oscillators
leads to the phase locking~(\ref{eq:SPhaseLocking}) on the
correlated time scales $s$.

When the parameter of coupling between chaotic oscillators
increases, more and more time scales become correlated and one can
say that the degree of the synchronization grows. So, with the
further increasing of the coupling parameter value (e.g.,
$\varepsilon=0.07$) in the coupled R\"ossler
systems~(\ref{eq:FunnelRoessler}) the time scales which were
uncorrelated before become synchronized (see
Fig.~\ref{fgr:FunRossK},\textit{c}). It is evident, that the time
scales $s=4.5$ are synchronized in comparison with the previous
case ($\varepsilon=0.05$, Fig.~\ref{fgr:FunRossK},\textit{b}) when
these time scales were uncorrelated. The number of time scales $s$
demonstrating the phase locking increases, but there are
nonsynchronized time scales as before (e.g., the time scales
$s=3.0$ and $s=6.0$ remain still nonsynchronized).

Arising of the lag
synchronization~\cite{Rosenblum:1997_LagSynchro} between
oscillators means that all time scales are correlated. Indeed,
from the condition of the lag--synchronization ${x_1(t-\tau)\simeq
x_2(t)}$ one can obtain that ${W_1(s,t-\tau)\simeq W_2(t,s)}$ and
therefore ${\phi_{s1}(t-\tau)\simeq\phi_{s2}(t)}$. It is clear, in
this case the phase locking condition~(\ref{eq:SPhaseLocking}) is
satisfied for all time scales $s$. For instance, when the coupling
parameter of chaotic oscillators~(\ref{eq:FunnelRoessler}) becomes
large enough ($s=0.25$) the lag synchronization of two coupled
oscillators appears. In this case the power spectra of wavelet
transform coincide with each other (see
Fig.~\ref{fgr:FunRossK},\textit{d}) and the phase locking takes
place for all time scales $s$
(Fig.~\ref{fgr:FunRossK},\textit{d}). It is important to note that
the phase difference $\phi_{s1}(t)-\phi_{s2}(t)$ is not equal to
zero for the case of the lag synchronization. It is clear that
this difference depends on the time lag $\tau$.

\begin{figure}[tb]
 \centerline{\includegraphics*[scale=0.4]{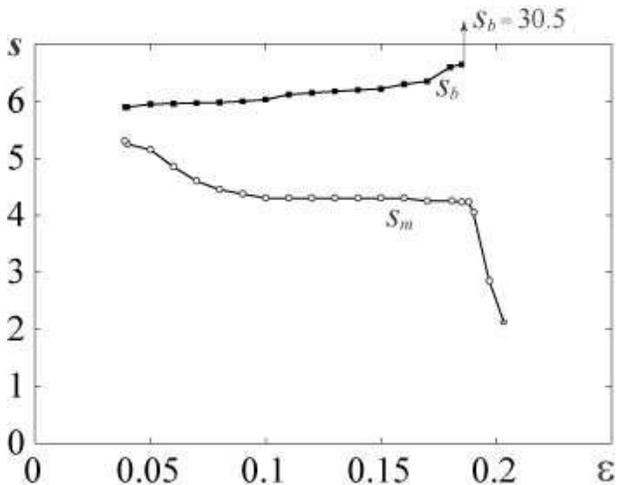}}
\caption{The dependence of the synchronized time scale range
$[s_m;s_b]$ on the coupling strength $\varepsilon$ for two
mutually coupled R\"ossler systems~(\ref{eq:FunnelRoessler}) with
funnel attractors\label{fgr:s}}
\end{figure}

Further increasing of the coupling parameter leads to the
decreasing of the time lag
$\tau$~\cite{Rosenblum:1997_LagSynchro}. Both systems tend to have
the complete synchronization regime $x_1(t)\simeq x_2(t)$, so the
phase difference $\phi_{s1}(t)-\phi_{s2}(t)$ tends to be a zero
for all time scales.

The dependence of synchronized time scale range $[s_m;s_b]$ on
coupling parameter $\varepsilon$ has been shown in
Fig.~\ref{fgr:s}. The range $[s_m;s_b]$ of synchronized time
scales appears at $\varepsilon\approx 0.039$. The appearance of
synchronized time scale range corresponds to the phase
synchronization regime. When the coupling parameter value
increases the range of synchronized time scales expands until all
time scales become synchronized. Synchronization of all time
scales means the presence of lag synchronization regime.

So, we can say the time scale synchronization (TSS) is the most
general synchronization type uniting (at least) PS, LS and CS
regimes. The regime of lag synchronization and the phase
synchronization differ from each other only in the number of
synchronized time scales.

\section{Example II. Time scale synchronization of Chua's circuits}
\label{Sct:Chua}

\begin{figure*}
 \centerline{\includegraphics*[scale=0.9]{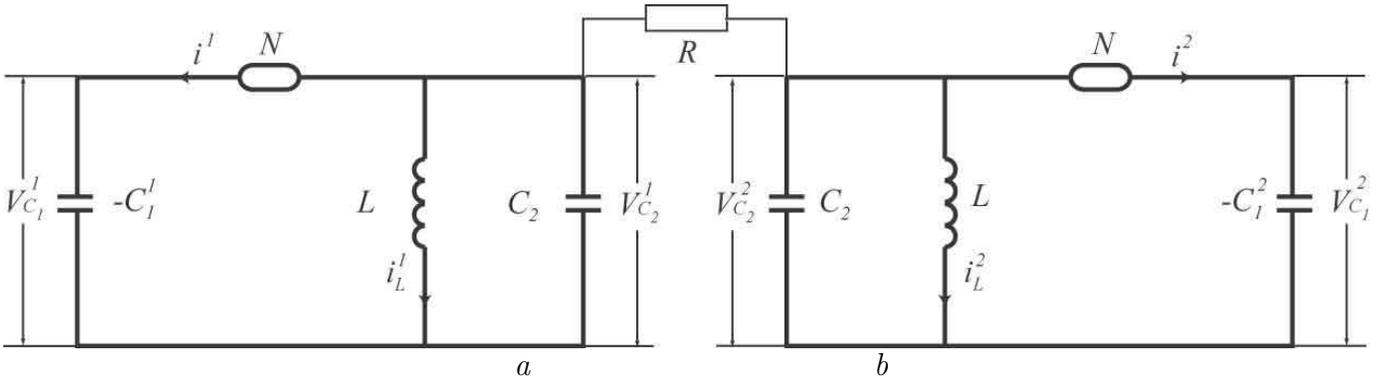}}
\centerline{\large\quad\textit{a}\qquad\qquad\qquad\qquad\qquad\quad\textit{b}}
\caption{Circuit realization of two mutually coupled Chua's
oscillators\label{fgr:Scheme}}
\end{figure*}

Let us consider the dynamics of two mutually coupled Chua's
circuits~\cite{Matsumoto:1987_TORUS,Chua:1992_Genesis} with
piece--wise linear characteristic (see Fig.~\ref{fgr:Scheme}). The
important feature of Chua's circuit is the presence of two
characteristic time scales $s_1$ and $s_2$ (or two characteristic
frequencies $f_1$ and $f_2$). It makes possible the realization in
this system both quasi--periodic and chaotic oscillations.

The behavior of two coupled oscillators is described by
\begin{equation}
\begin{array}{l}
\displaystyle
\dot x_{1,2}=-\frac{\alpha_{1,2}}{\gamma}f(y_{1,2}-x_{1,2}),\\
\displaystyle \dot
y_{1,2}=-\frac{1}{\gamma}\left(f(y_{1,2}-x_{1,2})+z_{1,2}\right)+
\frac{\varepsilon}{\gamma}(y_{2,1}-y_{1,2}),\\
\displaystyle\dot z_{1,2}=\gamma y_{1,2}, \label{eq:Torus}
\end{array}
\end{equation}
where $x_{1,2}=V_{C_1}^{1,2}/E$ and $y_{1,2}=V_{C_2}^{1,2}/E$ are
dimensionless voltages on capacitors $C_1^{1,2}$ and $C_2$ of the
first and the second oscillators, respectively. The variable
$z_{1,2}=i^{1,2}_L/I$ is the dimensionless current. The parameters
$E$ and $I$ are the normalization factors. Dimensionless control
parameters are $\alpha_{1,2}=C_2/C_1^{1,2}$ and
$\gamma=\frac{1}{m_1}\sqrt{{C_2}/{L}}$; $\tau=t/\sqrt{LC_2}$ is
dimensionless time. The function
\begin{equation}
f(\xi)=-\frac{m_0}{m_1}\xi+\frac{1}{2}
\left(\frac{m_0}{m_1}\right)\left(|\xi+1|-|\xi-1|\right),
\end{equation}
is the dimensionless voltage-current characteristic of nonlinear
element $N$, where $m_0$ and $m_1$ are the conductivities of the
corresponding branches of voltage-current characteristic. The
coupling parameter ${\displaystyle\varepsilon={1}/{Rm_1}}$
determines the influence of coupled Chua's circuits on each other.

\begin{figure}
 \centerline{\includegraphics*[scale=0.4]{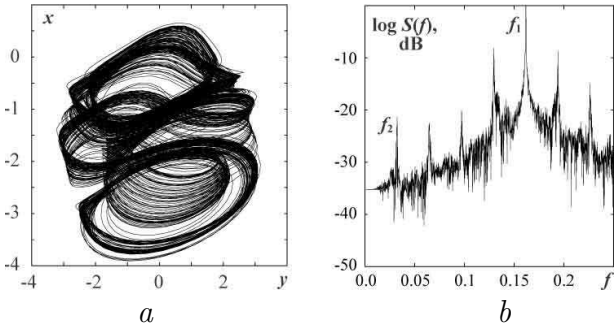}}
\centerline{\large\quad\textit{a}\qquad\qquad\qquad\qquad\qquad\quad\textit{b}}
\caption{The chaotic attractor and fourier spectrum of the first
Chua's circuit. The coupling parameter $\varepsilon$ is equal to
zero\label{fgr:Torus_spectrum}}
\end{figure}

The control parameter values have been selected as
$\alpha_1=2.78$, $\alpha_2=2.89$ and $\gamma=3.00$. The chaotic
attractor and fourier spectrum of the first Chua's circuit
oscillations is shown in Fig.~\ref{fgr:Torus_spectrum}, the
characteristic frequencies have been denoted as $f_1\simeq 0.161$
and $f_2\simeq 0.032$. The dynamics of the second Chua's circuit
is quite similar. Therefore, one can see in wavelet power spectra
$E_{1,2}(s)$ two maxima on time scales $s_1\simeq 6.2$ and
$s_2\simeq 30.0$ corresponding to the frequencies $f_1$ and $f_2$,
respectively.

\begin{figure}
 \centerline{\includegraphics*[scale=0.4]{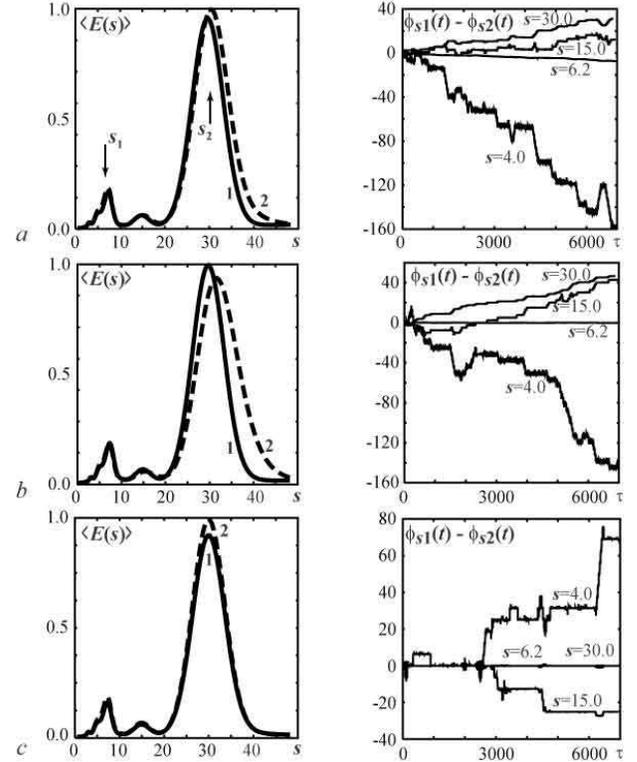}}
\caption{Wavelet power spectra of time series $x_{1,2}(t)$ of two
coupled Chua's circuit~(\ref{eq:Torus}) (solid line ``1''
corresponds to the first Chua's circuit, dashed line corresponds
to the second circuit, respectively) and phase difference
${\phi_{s1}(t)-\phi_{s2}(t)}$ on different time scales $s$. The
value of coupling parameter $\varepsilon$ is (\textit{a})
$\varepsilon=0.0$, (\textit{b}) $\varepsilon=0.05$, (\textit{c})
$\varepsilon=0.25$\label{fgr:TorusTSS}}
\end{figure}

When the coupling parameter is small TSS is not observed at all
(fig.~\ref{fgr:TorusTSS},\textit{a}). When the parameter
$\varepsilon$ increases the TSS appears on the time scales $s_1$
and close to them (see fig.~\ref{fgr:TorusTSS},\textit{b}).
Synchronous behavior on these time scales may be also detected as
phase synchronization of coupled chaotic systems~(\ref{eq:Torus})
by means of traditional approaches discussed
in~\cite{Pikovsky:2000_SynchroReview,Anishchenko:2002_SynchroEng,%
Pikovsky:2002_SynhroBook,Anshchenko:2001_SynhroBook,%
Rosenblum:1996_PhaseSynchro,Osipov:1997_PhaseSynchro}. With
further coupling parameter increasing the second time scales $s_2$
become also synchronized. For these time scales the phase locking
condition~(\ref{eq:SPhaseLocking}) and condition for wavelet
spectra energy~(\ref{eq:SynchroEnergy}) are satisfied (see
~\ref{fgr:TorusTSS},\textit{c}). It is important to note, that the
appearance of synchronized behavior on time scales $s_2$ (and
close to them) can not be detected by means of traditional
approaches as easily as before in the case of synchronous behavior
on time scales $s_1$. In this case the synchronization on time
scales $s_2$ is masked by synchronous behavior on time scales
$s_1$.

So, the TSS allows to analyse the chaotic behavior of the coupled
systems with several spectral basic components in fourier
spectrum. It is important to note that the synchronization
phenomena can take place on the several different time scale
ranges. In this case the energy being fallen on the synchronous
time scales should be calculated as
\begin{equation}
E_{snhr}=\int\limits_{s_{1m}}^{s_{1b}}\langle E(s)\rangle\,ds +
\int\limits_{s_{2m}}^{s_{2b}}\langle E(s)\rangle\,ds.
\label{eq:SynchroEnergy_2}
\end{equation}

\section{Example III. Generalized synchronization versus TSS}
\label{Sct:GSSynchro}

Let us consider now another type of synchronized behavior,
so--called the generalized synchronization. It has been shown
above, that PS, LS and CS are naturally interrelated with each
other and the synchronization type depends on the number of
synchronized time scales. The details of the relations between PS
and GS is not at all clear. There are several
works~\cite{Parlitz:1996_PhaseSynchroExperimental,%
Zhigang:2000_GSversusPS} dealing with the problem, how GS and PS
are correlated with each other. For instance,
in~\cite{Zhigang:2000_GSversusPS} it has been reported that two
unidirectional coupled R\"ossler systems can demonstrate the
generalized synchronization while the phase synchronization has
not been observed. This case allows to be explained easily by
means of the time scale analysis. The equations of R\"ossler
system are
\begin{equation}
\begin{array}{l}
\dot x_{1}=-\omega_{1}y_{1}-z_{1},\\
\dot y_{1}=\omega_{1}x_{1}+ay_{1},\\
\dot z_{1}=p+z_{1}(x_{1}-c)\\
\dot x_{2}=-\omega_{2}y_{2}-z_{2}+\varepsilon(x_1-x_2),\\
\dot y_{2}=\omega_{2}x_{2}+ay_{2},\\
\dot z_{2}=p+z_{2}(x_{2}-c), \label{eq:UniDirecRosslers}
\end{array}
\end{equation}
where $\mathbf{x}_1=(x_1,y_1,z_1)^T$ and
$\mathbf{x}_2=(x_2,y_2,z_2)^T$ are the state vectors of the first
(drive) and the second (response) R\"ossler systems, respectively.
The control parameter values have been chosen as $\omega_1=0.8$,
$\omega_2=1.0$, $a=0.15$, $p=0.2$, $c=10$ and $\varepsilon=0.2$.
The generalized synchronization takes place in this case
(see~\cite{Zhigang:2000_GSversusPS} for detail). Why it is
impossible to detect the phase synchronization in the
system~(\ref{eq:UniDirecRosslers}) despite the generalized
synchronization is observed becomes clear from the time scale
analysis.

\begin{figure}
 \centerline{\includegraphics*[scale=0.4]{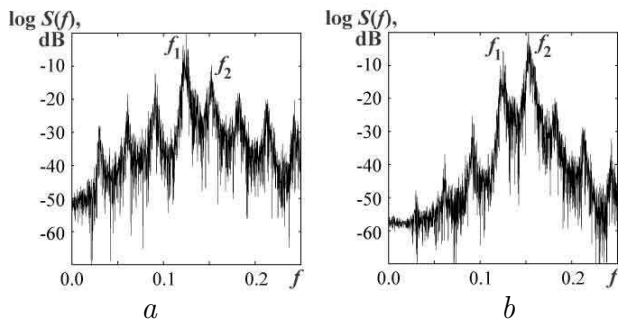}}
\centerline{\large\quad\textit{a}\qquad\qquad\qquad\qquad\qquad\quad\textit{b}}
\caption{Fourier spectra for(\textit{a}) the first (drive) and
(\textit{b}) the second (response) R\"osler
systems~(\ref{eq:UniDirecRosslers}). The coupling parameter is
$\varepsilon=0.2$. The generalized synchronization takes place
\label{fgr:GS2_spectra}}
\end{figure}

\begin{figure}
 \centerline{\includegraphics*[scale=0.4]{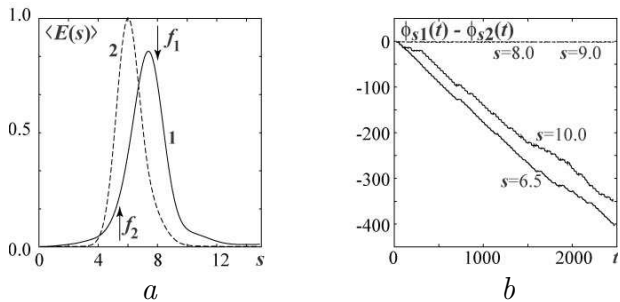}}
\centerline{\large\quad\textit{a}\qquad\qquad\qquad\qquad\qquad\quad\textit{b}}
\caption{(\textit{a}) The normalized energy distribution in
wavelet spectrum $\langle E(s)\rangle$ for the first (the solid
line denoted ``1'') and the second (the dashed line denoted ``2'')
R\"ossler systems. The time scales pointed with arrows correspond
to the frequencies $f_1=0.125$ and $f_2=0.154$, respectively;
(\textit{b}) the phase difference $\phi_{s1}(t)-\phi_{s2}(t)$ for
two coupled R\"ossler systems. The generalized synchronization has
been observed \label{fgr:GS2_wvt}}
\end{figure}

Let us consider Fourier spectra of coupled chaotic oscillators
(see Fig.~\ref{fgr:GS2_spectra}). There are two main spectral
components with frequencies $f_1=0.125$ and $f_2=0.154$ in these
spectra. The analysis of behavior of time scales has shown that
both the time scales $s_1=1/f_1=8$ of coupled oscillators
corresponding to the frequency $f_1$ and time scales close to
$s_1$ are synchronized while the time scales $s_2=1/f_2\simeq 6.5$
and close to them do not demonstrate synchronous behavior
(Fig.~\ref{fgr:GS2_wvt},\textit{b}).

The source of such behavior of time scales becomes clear from the
wavelet power spectra $\langle E(s)\rangle$ of both systems (see
Fig.~\ref{fgr:GS2_wvt},\textit{a}). The time scale $s_1$ of the
drive R\"ossler system is characterized by the large value of
energy while the part of energy being fallen on this scale of the
response system is quite small. Therefore, the drive system
dictates its own dynamics on the time scale $s_1$ to the response
system. The opposite situation takes place for the time scales
$s_2$ (see Fig.~\ref{fgr:GS2_wvt},\textit{a}). The drive system
can not dictate its dynamics to the response system because the
part of energy being fallen on this time scale is small in the
first R\"ossler system and large enough in the second one. So,
time scales $s_2$ are not synchronized.

Thus, the generalized synchronization of the unidirectional
coupled R\"ossler systems appears as the time scale synchronized
dynamics, as another synchronization types does before. It is also
clear, why the phase synchronization has not been observed in this
case. The instantaneous phases $\phi_{1,2}(t)$ of chaotic signals
$\mathbf{x}_{1,2}(t)$ introduced by means of traditional
approaches are determined by both frequencies $f_1$ and $f_2$, but
only the spectral components with the frequency $f_1$ are
synchronized. So, the observation of instantaneous phases
$\phi_{1,2}(t)$ does not allow to detect the phase synchronization
in this case although the synchronization of time scales takes
place.

\begin{figure}
 \centerline{\includegraphics*[scale=0.4]{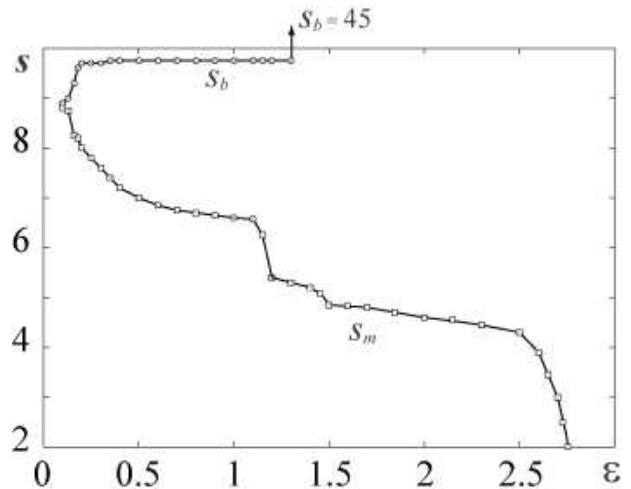}}
\caption{The dependence of the synchronized time scale range
$[s_m;s_b]$ on the coupling strength $\varepsilon$ for two
unidirectionally coupled R\"ossler
systems~(\ref{eq:UniDirecRosslers})\label{fgr:s1}}
\end{figure}

With increasing the coupling parameter between systems the range
of synchronized time scales increases (Fig.~\ref{fgr:s1}) as well
as in the case of phase synchronization (see
sec.~\ref{Sct:IllPhase}). When all time scales become synchronized
the lag synchronization appears.

Thus, one can see that there is a close relationship between
different types of the chaotic oscillator synchronization.
According to results mentioned above we can say that PS, LS, CS
and GS are particular cases of TSS. Therefore, it is possible to
consider different types of synchronized behavior from the
universal position. Unfortunately, it is not clear, how one can
distinguish the phase synchronization and the generalized
synchronization using only the results obtained from the analysis
of the time scale dynamics. (Here we mean the phase
synchronization between chaotic oscillators takes place if the
instantaneous phase $\phi(t)$ of chaotic signal may be correctly
introduced by means of traditional approaches and the phase
locking condition~(\ref{eq:PhaseLocking}) is satisfied.)

\section{Measure of synchronization}
\label{Sct:Measure}

From examples mentioned above one can see that any type of
synchronous behavior of coupled chaotic oscillators leads to
arising of the synchronized time scales. Therefore, the measure of
synchronization can be introduced. This measure $\rho$ can be
defined as the the part of wavelet spectrum energy being fallen on
the synchronized time scales
\begin{equation}
\rho_{1,2}=\frac{1}{E_{1,2}}\int\limits_{s_m}^{s_b}\langle
E_{1,2}(s)\rangle\,ds,
\end{equation}
where $[s_m;s_b]$ is the range of time scales for which the
condition~(\ref{eq:PhaseLocking}) is satisfied and $E_{1,2}$ is a
full energy of the wavelet spectrum
\begin{equation}
E_{1,2}=\int\limits_{0}^{+\infty}\langle E_{1,2}(s)\rangle\,ds.
\end{equation}
This measure $\rho$ is 0 for the nonsynchronized oscillations and
1 for the case of complete and lag synchronization regimes. If the
phase synchronization regime is observed it takes a value between
0 and 1 depending on the part of energy being fallen on the
synchronized time scales. So, the synchronization measure $\rho$
allows not only to distinguish the synchronized and
nonsynchronized oscillations, but characterize quantitatively the
degree of TSS synchronization.

\begin{figure}
 \centerline{\includegraphics*[scale=0.4]{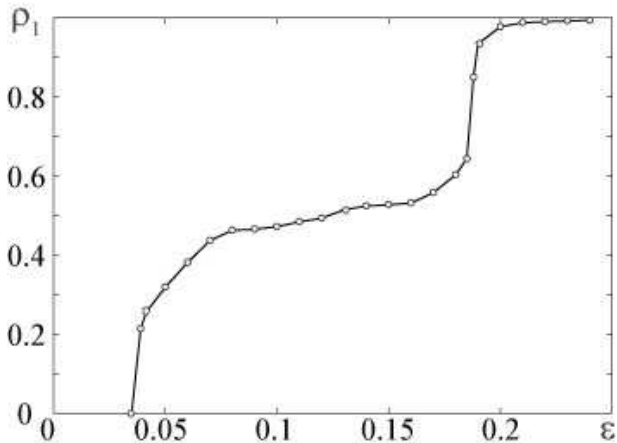}}
\caption{The dependence of the synchronization measure $\rho_1$
for the first R\"ossler system~(\ref{eq:FunnelRoessler}) on the
coupling strength $\varepsilon$. The measure $\rho_2$ for the
second R\"ossler oscillator behaves in a similar manner, so it has
not been shown in the figure\label{fgr:rho}}
\end{figure}

\begin{figure}
 \centerline{\includegraphics*[scale=0.4]{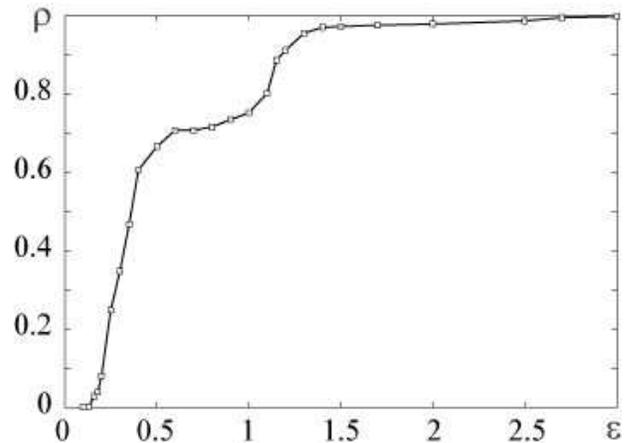}}
\caption{The dependence of the synchronization measure $\rho$ for
the second R\"ossler system~(\ref{eq:UniDirecRosslers}) on the
coupling strength $\varepsilon$\label{fgr:g1}}
\end{figure}

Fig.~\ref{fgr:rho} presents the dependence of the TSS
synchronization measure $\rho_1$ for the first R\"ossler
oscillator of system~(\ref{eq:FunnelRoessler}) considered in
section~\ref{Sct:IllPhase} on the coupling parameter
$\varepsilon$. It is clear that the part of the energy being
fallen on the synchronized time scales growths monotonically with
the growth of the coupling strength. Similar results
(Fig.~\ref{fgr:g1}) have been obtained for the generalized
synchronization of two coupled R\"ossler systems considered in
section~\ref{Sct:GSSynchro}.

It has already mentioned above that when the coupled oscillators
do not demonstrate synchronous behavior there are time scales
$s^*$ the phase difference $\phi_{s1}(t)-\phi_{s2}(t)$ on which is
bounded. Such time scales play role of points separating the time
scale areas where the phase difference increases and decreases,
respectively (see also section~\ref{Sct:IllPhase}). Nevertheless,
the presence of such time scales does not mean the occurrence of
chaotic synchronization because the part of energy being fallen on
them is equal to zero. Therefore, the synchronization measure
$\rho$ of such oscillations is zero, and dynamical regime being
realized in the system in this case should be classified as
non-synchronous.

\section{Conclusion}
\label{Sct:Conclusion}

Summarizing this work we would like to note several principal
aspects. Firstly, we have proposed to consider the time scale
dynamics of coupled chaotic oscillators. It allows us to consider
the different types of behavior of coupled oscillators (such as
the complete synchronization, the lag synchronization, the phase
synchronization, the generalized synchronization and the
nonsynchronized oscillations) from the universal position. In this
case TSS is the most common type of synchronous coupled chaotic
oscillator behavior. Therefore, the another types of synchronous
oscillations (PS, LS, CS and GS) may be considered as the
particular cases of TSS. The quantitative characteristic $\rho$ of
the synchronization measure has also been introduced. It is
important to note that our method (with insignificant
modifications) can also be applied to dynamical systems
synchronized by the external (e.g., harmonic) signal.

Secondly, the traditional approach for the phase synchronization
detecting based on the introducing of the instantaneous phase
$\phi(t)$ of chaotic signal is suitable and correct for such time
series characterized by the Fourier spectrum with the single main
frequency $f_0$. In this case the phase $\phi_{s0}$ associated
with the time scale $s_0$ corresponding to the main frequency
$f_0$ of the Fourier spectrum coincides approximately with the
instantaneous phase $\phi(t)$ of chaotic signal introduced by
means of the traditional approaches (see
also~\cite{Quiroga:2002_Kraskov}). Indeed, as the other
frequencies (the other time scales) do not play a significant role
in the Fourier spectrum, the phase $\phi(t)$ of chaotic signal is
close to the phase $\phi_{s0}(t)$ of the main spectral frequency
$f_0$ (and the main time scale $s_0$, respectively). It is
obvious, that in this case the mean frequencies
$\bar{f}=\langle\dot{\phi}(t)\rangle/2\pi$ and
$\bar{f}_{s0}=\langle\dot{\phi}_{s0}(t)\rangle/2\pi$ should
coincide with each other and with the main frequency $f_0$ of the
Fourier spectrum (see also~\cite{Anishchenko:2004_ChaosSynchro})
\begin{equation}
\bar{f}=\bar{f}_{s0}=f_0.
\end{equation}
If the chaotic time series is characterized by the Fourier
spectrum without the main single frequency (like the spectrum
shown in the Fig.~\ref{fgr:FunnelRoessler},\textit{b}) the
traditional approaches fail. One has to consider the dynamics of
the system on all time scales, but it is impossible to do it by
means of the instantaneous phase $\phi(t)$. On the contrary, our
approach based on the time scale dynamics analysis can be used for
both types of chaotic signals.

Finally, our approach can be easily applied to the experimental
data because it does not require any \hbox{a-priori} information
about the considered dynamical systems. Moreover, in several cases
the influence of the noise can be reduced by means of the wavelet
transform (for detail,
see~\cite{alkor:2003_WVTBookEng,Torrence:1998_Wvt,Gusev:2003_wvt}).
We believe that our approach will be useful and effective for the
analysis of physical, biological, physiological and other data,
such as~\cite{Elson:1998_NeronSynchro,Quiroga:2002_Kraskov,%
Lachaux:2000_WVTSynchro}.

\section*{Acknowledgments}
\label{Sct:Acknowledgments} We express our appreciation to George
A. Okrokvertskhov, Alexander V. Kraskov and Professors
Vadim~S.~Anishchenko and Tatyana E. Vadivasova for valuable
discussions. We also thank Svetlana V. Eremina for the support.

This work has been supported by U.S.~Civilian Research \&
Development Foundation for the Independent States of the Former
Soviet Union (CRDF), grant {REC--006}. A.E.H. also thanks
``Dynastiya'' Foundation.


\end{document}